\documentclass{article}
\usepackage{ijcai24}

\usepackage{times}
\usepackage{soul}
\usepackage{url}
\usepackage[hidelinks]{hyperref}
\usepackage[utf8]{inputenc}
\usepackage[small]{caption}
\usepackage{graphicx}
\usepackage{amsmath}
\usepackage{amsthm}
\usepackage{booktabs}
\usepackage{algorithm}
\usepackage{algorithmic}
\usepackage[switch]{lineno}

\usepackage{xcolor}
\usepackage{amssymb}

\newtheorem{theorem}{Theorem}

\title{Smuche: Scalar-Multiplicative Caching in Homomorphic Encryption}

\author{
    Dongfang Zhao
    \affiliations
    University of Washington
    \emails
    dzhao@uw.edu
}

\begin{document}

\maketitle

\begin{abstract}
Addressing the challenge of balancing security and efficiency when deploying machine learning systems in untrusted environments, such as federated learning, remains a critical concern. A promising strategy to tackle this issue involves optimizing the performance of fully homomorphic encryption (HE). Recent research highlights the efficacy of advanced caching techniques, such as Rache, in significantly enhancing the performance of HE schemes without compromising security. However, Rache is constrained by an inherent limitation: its performance overhead is heavily influenced by the characteristics of plaintext models, specifically exhibiting a caching time complexity of $\mathcal{O}(N)$, where $N$ represents the number of cached pivots based on specific radixes. This caching overhead becomes impractical for handling large-scale data.
In this study, we introduce a novel \textit{constant-time} caching technique that is independent of any parameters. The core concept involves applying scalar multiplication to a single cached ciphertext, followed by the introduction of a completely new and constant-time randomness. Leveraging the inherent characteristics of constant-time construction, we coin the term ``Smuche'' for this innovative caching technique, which stands for Scalar-multiplicative Caching of Homomorphic Encryption. We implemented Smuche from scratch and conducted comparative evaluations against two baseline schemes, Rache and CKKS. Our experimental results underscore the effectiveness of Smuche in addressing the identified limitations and optimizing the performance of homomorphic encryption in practical scenarios.
\end{abstract}

\section{Introduction}

\paragraph{Background}
Machine learning systems deployed in untrustworthy environments must implement security measures with acceptable performance overhead. This is evident in the context of Federated Learning (FL) systems~\cite{mcmahan_aistat17}, where participating clients avoid sharing their private data with the parameter server. Even the models trained on private data should remain confidential, as their exposure to the parameter server could potentially lead to the leakage of the training dataset~\cite{lzhu_nips19}. A direct solution to this concern is the adoption of (fully) homomorphic encryption (HE), encrypting local models before uploading them to the parameter server~\cite{tenseal}. However, HE cryptosystems address security concerns at the expense of introducing significant computational overhead, prompting recent efforts to optimize HE performance through advanced caching techniques~\cite{otawose_sigmod23}.

\paragraph{Motivation}

While the state-of-the-art optimization in homomorphic encryption (HE), specifically Rache~\cite{otawose_sigmod23}, demonstrates significant improvements over vanilla HE deployments, a notable limitation lies in the scalability of caching ciphertexts. More precisely, the computational overhead scales on the order of $\mathcal{O}(N)$, where $N$ represents the number of cached ciphertexts during encryption. This linear growth reveals scalability challenges, especially when $N$ becomes sufficiently large. Indeed, our experiments (see Table~\ref{tbl:rache_scale}) indicate that caching an excessive number of ciphertext messages can lead to an overall performance degradation, potentially worse than that of the vanilla HE scheme.
This concrete overhead arises from both the construction of ciphertexts and the randomization procedure, both of which rely on homomorphic addition operations among ciphertexts. Although the homomorphic addition operation is approximately 10---100$\times$ less expensive than the homomorphic encryption operation, an excessive number of such addition operations can defeat the purpose of additive caching.

\subsection{Proposed Approach}

This research aligns with the ongoing exploration of enhancing the performance of homomorphic encryption (HE) schemes without compromising their security guarantees. Our objective is to boost the efficiency of state-of-the-art HE schemes while upholding their security. Specifically, we aim to eliminate the linear costs associated with two components:
(i) The linear construction of a new ciphertext based on cached pivots and
(ii) The linear randomization of the newly constructed ciphertext.

We demonstrate that both components can be efficiently implemented in \textit{constant time}. Achieving constant-time caching requires delving into the kernel of HE schemes, rather than treating existing HE schemes (e.g., CKKS~\cite{ckks}) as black boxes, as done in the case of Rache~\cite{otawose_sigmod23}. The fundamental idea behind constant-time construction is leveraging the scalar-multiplicative property of HE schemes. We briefly outline the high-level intuition behind both constant-time procedures:

\begin{itemize}
    \item Constant-time construction: Essentially, any additive HE scheme can support scalar multiplication between a ciphertext and an integer, as this multiplication can be trivially expanded into a series of pairwise additions. The challenge lies in whether scalar multiplication can outperform a series of additions. As detailed in later sections, this depends on the number of additions needed to construct the ciphertext message.
    
    \item Constant-time randomization: Instead of manipulating the cached ciphertext as a black box, we directly modify the randomness part of the ciphertext, comprising two polynomials. This involves a meticulous decomposition of the polynomials and a discrete parameterization of the underlying HE scheme. Additionally, the success of this strategy depends on the efficiency of kernel-level randomization, i.e., whether it is more cost-effective than the black box solution found in the state-of-the-art.
\end{itemize}

\subsection{Contributions}

This paper presents the following contributions:

\begin{enumerate}
    \item We introduce Smuche, a constant-time caching technique designed to facilitate high-performance homomorphic encryption of large-scale data. Smuche aims to strike a balance between performance and security in secure machine learning, representing the first algorithm of its kind.

    \item We establish the provable security of Smuche. The proof follows the classical reduction framework by reducing the challenge of breaking Smuche to distinguishing a ciphertext encrypted by the underlying HE scheme and a random string, which is assumed to be computationally infeasible.

    \item We implement Smuche from the ground up and conduct a comparative analysis against baseline schemes Rache and CKKS. Experimental results affirm the effectiveness of Smuche, showcasing constant-time caching of encrypted ciphertext with significantly lower overhead than the state-of-the-art.
\end{enumerate}

\section{Preliminaries and Related Work}

\subsection{Homomorphic Encryption}
The notion of \textit{homomorphism} originates from the study of algebraic groups~\cite{fraleigh_book03}, which are algebraic structures defined over nonempty sets. Formally, a group $G$ over a set $S$ is represented as a tuple $(G, \oplus)$, where $\oplus$ is a binary operator. This operator adheres to four fundamental axioms or properties, expressed through first-order logical formulas:
(i) For all $g, h \in S$, $g \oplus h \in S$.
(ii) There exists a unique element $u \in S$ such that for all $g \in S$, $(g \oplus u = g)$ and $(u \oplus g = g)$.
(iii) For every $g \in S$, there exists an element $h \in S$ such that $(g \oplus h = u)$ and $(h \oplus g = u)$. This element $h$ is often denoted as $-g$.
(iv) For all $g, h, j \in S$, $(g \oplus h) \oplus j = g \oplus (h \oplus j)$.
Given another group $(H, \otimes)$ and a function $\varphi: G \rightarrow H$ that for all $g_1, g_2 \in G$ satisfies $\varphi(g_1) \otimes \varphi(g_2) = \varphi(g_1 \oplus g_2)$, we define $\varphi$ as a homomorphism from $G$ to $H$. This implies that $\varphi$ preserves the group operation between the elements of $G$ when mapped to the corresponding elements in $H$.

\textit{Homomorphic encryption} (HE) is a specific type of encryption where certain operations between operands can be performed directly on the ciphertexts.
For example, if an HE scheme $he(\cdot)$ is additive,
then the plaintexts with $+$ operations can be translated into a homomorphic addition $\oplus$ on the ciphertexts.
Formally, if $a$ and $b$ are plaintexts, the relationship is defined as:
\[
dec(he(a) \oplus he(b)) = a + b,
\]
where $dec$ denotes the decryption algorithm.
As a concrete example, setting $he(x) = 2^x$ (temporarily disregarding security considerations of $he(\cdot)$) demonstrates that $he(a+b) = 2^{a+b} = 2^a \times 2^b = he(a) \times he(b)$, indicating that $\oplus$ corresponds to arithmetic multiplication $\times$.

An HE scheme enabling additive operations is termed \textit{additive}.
Popular additive HE schemes include Paillier~\cite{ppail_eurocrypt99},
which is an asymmetric scheme using public and private keys for encryption and decryption, respectively. 
An HE scheme that supports multiplication is said to be \textit{multiplicative}.
Symmetria~\cite{symmetria_vldb20} is a recent scheme proposed in the database community,
which is multiplicative using a distinct scheme from the one for addition.
Other well-known multiplicative HE schemes include RSA~\cite{rsa} and ElGamal~\cite{elgamal_tit85}.
Similarly, a multiplicative HE scheme guarantees the following equality,
\[
dec (he(a) \otimes he(b)) = a \times b,
\]
where $\otimes$ denotes the homomorphic multiplication over the ciphertexts.

An HE scheme that supports both addition and multiplication is classified a \textit{fully HE (FHE) scheme}.
This requirement should not be confused with specific addition and multiplication parameters, such as Symmetria~\cite{symmetria_vldb20} and NTRU~\cite{ntru}.
That is, the addition and multiplication must be supported homomorphically under the same scheme $he(\cdot)$:
\[\displaystyle
\begin{cases}
    dec( he(a) \oplus he(b) ) = a + b \\
    dec( he(a) \otimes he(b) ) = a \times b.
\end{cases}
\]
Constructing FHE schemes remained a formidable challenge until Gentry~\cite{cgentry_stoc09} presented a feasible approach using lattice ideals.
Although lattice has been extensively studied in cryptography, the combination with ring ideals was less explored; nonetheless, Gentry showed that it is possible to construct an FHE scheme although the cost to maintain the multiplicative homomorphism is prohibitively significant even with the so-called bootstrapping optimization,
which essentially applies decryption for every single multiplication operation between ciphertexts.

Subsequent generations of FHE schemes,
e.g., \cite{bv11,bgv,bfv,ckks}, have seen substantial improvements in encryption efficiency, partially due to the removal of ideal lattices;
rather the new series of FHE schemes are based on the learning with error (LWE)~\cite{oregev_jacm09} or its variant ring learning with error (RLWE),
which have been proven to be as secure as hard lattice problems (e.g., through quantum or classical reduction).
The positive development is that schemes that are built upon LWE or RLWE are significantly more efficient than the first-generation schemes.
However, a significant performance gap remains between the current FHE cryptosystems and the desired efficiency levels. Open-source libraries of FHE schemes, such as IBM HElib~\cite{helib} and Microsoft SEAL~\cite{sealcrypto}, are available and represent the state-of-the-art in the field.

\subsection{Threat Model and Provable Security}

Informally, a threat model portraits the capabilities of the attackers.
One widely used model is chosen-plaintext attack (CPA), where the adversary is assumed to have the ability to obtain the ciphertext corresponding to any plaintext of their choosing.
In practice, it is assumed that the adversary has limitations on the number of plaintext-ciphertext pairs they can obtain. Specifically, the access of adversary should be restricted to only a polynomial number of such pairs, and their computational resources are assumed to be bounded by polynomial time. This assumption reflects the notion that adversaries should not possess unlimited computational power or the ability to amass an impractical volume of information.
By making these assumptions, security analysis can be conducted under realistic constraints, allowing for a comprehensive evaluation of the encryption scheme's resilience against chosen-plaintext attacks. The goal is to design an encryption scheme that maintain its security even when confronted with an adversary who can strategically chooses plaintexts and acquires their corresponding ciphertexts within the given limitations.

In the ideal scenario, even if the adversary $\mathcal{A}$ successfully obtain additional information, $\mathcal{A}$ should not be able to make a \textit{distinguishably} better decision for the plaintexts than what would be achievable through random guess. This principle is encapsulated in the concept of a \textit{negligible function}.
A function $\mu(\cdot)$ is defined negligible when, for all polynomials $poly(n)$, the inequality $\mu(n) < \frac{1}{poly(n)}$ holds for sufficiently large values of $n$. 
In other words, the function $\mu(\cdot)$ decreases faster than the reciprocal of any polynomial as $n$ grows. The use of negligible functions is critical in quantifying the extent of the advantage an adversary could gain over random guessing. They are crucial component in the verification of security proofs within cryptographic systems.

The security of cryptographic schemes, especially homomorphic encryption, is often assessed through formal proofs based on established security notions. The Indistinguishability under Chosen Plaintext Attack (IND-CPA) security is a fundamental concept that assures the confidentiality of encrypted data even when adversaries have access to chosen plaintext-ciphertext pairs. The methodology for providing IND-CPA security incorporates reduction techniques, analysis of probabilities, and the application of negligible functions.

The de facto approach to prove IND-CPA security is through reduction.
The reduction approach in IND-CPA proofs template involves relating the security of a cryptographic scheme to the assumed infeasibility of a particular computational problem. Commonly, this involves the challenges of breaking the underlying cryptographic primitive. In the context of homomorphic encryption, this reduction demonstrates that if an adversary is capable to distinguish between two ciphertexts encrypted from distinct plaintexts, then it implies that they could also solve the assumed computational problem, which is considered inherently difficult.

At the core of proof template are negligible functions, which are defined by their property that decrease faster than the reciprocal of any polynomial. In the context of security proofs, if the advantage of an adversary (the probability that they can distinguish between ciphertexts) is bounded by a negligible function, then the scheme is considered IND-CPA secure. This notion encapsulates the concept that as the input size grows, the negligible function diminishes more rapidly than any polynomial increase.

\subsection{Rache: Radix-based Additive Caching}

Rache~\cite{otawose_sigmod23} represents an optimization of the expensive operation of homomorphic encryption by calculating the desired ciphertext from existing (i.e., cached) ciphertexts.
Rache comprises three stages: precomputing, construction, and randomization. We briefly review them in the following. 

\paragraph{Precomputing}
In this stage, the scheme selects a radix value $r$, calculates the integral exponentiation of $r$, also known as \textit{pivots}, usually up to the maximal number of the plaintext space, and encrypts them in a cache. The most straightforward choice of $r$ is 2, although other values are possible.\footnote{Rache picks a heuristic value of $r = 6$.}

\paragraph{Construction}
The ciphertext of a new message $m$, denoted by $enc(m)$, can be constructed from pivots $c_i$'s:
\[\displaystyle
c' \stackrel{def}{=} enc(m) = \bigoplus_{i=0}^{\lceil \log m \rceil - 1} (m \;\text{mod}\; r^{i+1}) \odot c_i,
\]
such that 
\[\displaystyle
m = \sum_{i=0}^{\lceil \log m \rceil - 1} (m \;\text{mod}\; r^{i+1}) \cdot r^i,
\]
where $c_i$ is the $i$-th pivot in the cache and $\odot$ denotes the multiplication between an integer and a ciphertext.

\paragraph{Randomization}
The final randomized ciphertext of $m$ is calculated as
\[\displaystyle
c \stackrel{def}{=} c' \bigoplus_{i=0}^{\lceil \log m \rceil - 1} rnd_i \odot (c_{i+1} \ominus r \odot c_i),
\]
where $rnd_i$ denotes a random boolean value uniformly sampled from $\{0, 1\}$ and $\ominus$ denotes the minus operation in the ciphertext space. 

\section{Scalar-Multiplicative Caching}

The proposed Smuche caching scheme also comprises three stages:
precomputing, construction, and randomization. As we will see, each of these three stages adopts a completely different algorithm than that of Rache.

\subsection{Precomputing}

Smuche reuses the term \textit{pivots} that are precomputed ciphertexts of ``interesting'' plaintext values.
However, as opposed to the logarithmic intermediates adopted by Rache, Smuche chooses to cache those unit values at various decimal positions. By unit values, we mean the special invariants whose multiplications over any other value equal that value up to a scale. Such a unit is also called the \textit{multiplicative identity} in an algebraic ring structure. 

The most intuitive identity is number one for the integers $\mathbb{Z}$. However, float numbers could not be trivially reconstructed from one due to the fractional part. This issue can be solved by employing the scaling idea of modern HE schemes. Essentially, we introduce a scale $\Delta$ that can convert a float into an integer. For decimals, $\Delta$ can be simply set to $10^i$, $i \in \mathbb{Z}^+$. Therefore, a float number such as $1.02$ can be coded as $(z,\Delta) = (102,10^2)$. Consequently, we can set the pivot as $enc(\Delta^{-1})$ and any float number with the same decimal precision can be expressed in terms of $enc(\Delta^{-1})$.

\begin{algorithm}[tb]
    \caption{Smuche Precomputing}
    \label{alg:precompute}
    \textbf{Input} $\Delta^{-1}$: highest precision in the plaintext space; $r$: radix;\\
    \textbf{Output} $cache[\;]$: array of cached ciphertexts \\
    \begin{algorithmic}[1] 
        \STATE $i = 0$
        \FOR{$pivot = 1;\; pivot \ge \Delta^{-1};\; pivot = pivot/r$}
            \STATE $cache[i] = enc(r^{-i})$
            \STATE $i = i + 1$
        \ENDFOR
    \end{algorithmic}
\end{algorithm}

We present the precomputing procedure in Algorithm~\ref{alg:precompute}. Smuche caches all the exponents of the highest precision $\Delta^{-1}$ until integer one, which is the multiplicative identity of the integer ring. The overall number of cached ciphertexts depends on not only the highest precision but also the radix $r$, which serves as the step size in the precomputing. 

The precomputing procedure is usually carried out at an offline stage. Therefore, the cost is usually not a concern in practice. Nonetheless, even the asymptotic complexity of the precomputing algorithm is fairly efficient: $\mathcal{O}(\log \Delta)$, which is logarithmic in the precision of the plaintexts and does not depend on the scales of the plaintexts. In other words, this cost is constant regarding the plaintext size. This is in contrast to the offline overhead of Rache, which is logarithmic in the maximal value of the plaintext space.

\subsection{Construction}

The construction procedure is straightforward: A pivot is selected according to the precision of the given plaintext, after which the ciphertext is constructed through a scalar multiplication between the pivot and the scaled value of the plaintext. We stress that the scalar multiplication can be implemented as a series of ciphertext additions and should not require a homomorphic multiplication directly over a pair of ciphertexts.

\begin{algorithm}[tb]
    \caption{Smuche Construction}
    \label{alg:construct}
    \textbf{Input} $m$: plaintext message; $\delta^{-1}$: precision of $m$; $cache[\;]$: array of cached ciphertexts; $r$: radix\\
    \textbf{Output} $c'$: ciphertext such that $dec(c) = m$\\
    \begin{algorithmic}[1] 
        \STATE $idx = 0$
        \WHILE{$r^{-idx} > \delta^{-1}$}
            \STATE $idx = idx + 1$
        \ENDWHILE
        \STATE $c' = cache[idx] \odot (m \cdot \delta)$
    \end{algorithmic}
\end{algorithm}

The construction procedure is described in Algorithm~\ref{alg:construct}. Lines 1--4 search for the largest pivot in the cache such that a scalar multiplication in Line 5 could construct the desired ciphertext. The reason for picking the largest pivot is to reduce the scalar coefficient of the multiplicative construction such that the noise level is the lowest possible.

The complexity of the construction algorithm is similar to that of the precomputing. It should be noted that $|\delta| \le |\Delta|$. The loop on Lines 2--4 takes up to $\lceil \log_r \delta \rceil$ steps, which are asymptotically $\mathcal{O}(\log \Delta)$. Similarly to the procomputing algorithm, this cost is independent of the plaintext size and asymptotically constant.

\subsection{Randomization}

To fully understand the proposed randomization in Smuche, it is necessary to clarify the ciphertext structure of modern HE schemes, such as BFV~\cite{bfv}, BGV~\cite{bgv}, and CKKS~\cite{ckks}. Such a low-level detail is unnecessary for Rache because the caching algorithms in Rache deal with only the public primitives (i.e., programming interface) exposed by the underlying HE schemes. On the other hand, we will not be able to cover all the details of the ciphertext format due to the limited space; we refer interested readers to the original papers of those HE schemes. Instead, in what follows we will provide a high-level discussion.

A ciphertext message in HE schemes consists of two polynomials in the format of $enc(m) = c = (c_1, c_2)$, where $c_1, c_2 \in \mathbb{Z}_q/(X^N + 1)$, which denotes a set of polynomials with integer coefficients modulo $q$ and degrees modulo $N$.\footnote{More formally, it is a quotient ring over irreducible polynomial ring $X^N + 1$.}
The polynomial $c_2$ is an auxiliary piece of information to help decrypt the message that is masked in the polynomial $c_1$.
Both $c_1$ and $c_2$ are constructed with the public key $PK$, which is also defined as a pair of polynomials $PK = (PK_1, PK_2)$. To understand the full picture, we first define the private key $SK$ as follows:
\begin{equation}
\displaystyle
\label{eq:ckks_sk}
SK \gets R_2,
\end{equation}
where $R_2$ denotes a polynomial of degree $N$ with coefficient in $\{-1, 0, +1\}$ and $\gets$ denotes an independent and uniform sampling from the right-hand-side set. Informally, we pick our private key as a ``small'' polynomial in the sense that the $||\cdot||_{\infty}$ norm of coordinate-wise coefficients is less than 2. The public key $PK$ is then defined as follows:
\begin{equation}
\displaystyle
\label{eq:ckks_pk}
\begin{cases}
    PK_1 &=_{q} -a \cdot SK + e \\
    PK_2 &=_{q} a \gets R_q
\end{cases}
\end{equation}
where $=_q$ denotes the modulo $q$ operation for the assignment, $R_q$ is a polynomial whose coefficient is modulo $q$, and $e$ is a polynomial whose coefficients are sampled from a discrete Gaussian distribution with $\sigma = 3.2$. Informally, $e$ is a ``small'' polynomial while $a$ is a ``large'' polynomial in the ciphertext space.
We are now ready to encrypt a plaintext $m$.\footnote{There is usually an encoding procedure to convert a given scalar number (bit string) into a polynomial, but Smuche does not touch on that part so we will skip that discussion.}
Let $u \gets R_2$ and $e_1, e_2$ are sampled just as $e$ in the above (that is, all of these three polynomials are ``small''), the ciphertext $c = (c_1, c_2)$ is defined as
\begin{equation}
\displaystyle
\label{eq:ckks_ctxt}
\begin{cases}
    c_1 =_q PK_1\cdot u + e_1 + m\\
    c_2 =_q PK_2\cdot u + e_2
\end{cases}   
\end{equation}
The decryption algorithm $dec$ is defined as
\begin{equation}
\displaystyle
\label{eq:ckks_dec}
dec(c) = (c_1 + c_2 \cdot SK)_{\text{mod }q}.
\end{equation}
We skip the correctness of the decryption algorithm, which can be verified through elementary algebra and ignoring those ``small'' polynomial terms.

The addition in the ciphertext space, i.e., $\oplus$, is defined as the element-wise addition of the two polynomials in the ciphertext.
That is, the addition of two ciphertexts $c^1$ and $c^2$ is:
\begin{equation}
\displaystyle
\label{eq:ckks_add}
c^1 \oplus c^2 = \left( (c^1_1 + c^2_1)_{\text{mod }q} , (c^1_2 + c^2_2)_{\text{mod }q}\right),
\end{equation}
where $(\cdot)_{\text{mod }q}$ denotes the modulo $q$ of the value in the paratenses. 
From Eq.~\eqref{eq:ckks_add}, it is not hard to see that we can calculate the scalar multiplication of a ciphertext $c$ as
\begin{equation}
\displaystyle
\label{eq:ckks_smult}
z\odot c = \left( (z\cdot c_1)_{\text{mod }q} , (z\cdot c_2)_{\text{mod }q}\right),
\end{equation}
where $z \in \mathbb{Z}$.
Indeed, Line 5 of Alg.~\ref{alg:construct} does exactly that.
Now the question is, how can we make Eq.~\eqref{eq:ckks_smult} randomized\footnote{A randomized ciphertext is a hard requirement for any encryption scheme to be practically secure, i.e., semantically secure or IND-CPA (more on this in the next section).},
and that is what we will do in the remainder of this section.

Let $\xi \gets R_2$. We calculate the following random tuple 
\begin{equation}
\displaystyle
\label{eq:smuche_rnd}
rnd = \left( (PK_1\cdot \xi)_{\text{mod }q}, (PK_2\cdot \xi)_{\text{mod }q} \right).
\end{equation}
We claim that $z \odot c \oplus rnd$ is a valid encryption of $z\cdot m$.
That is, we will prove that the following equation holds.
\begin{theorem}
$dec(z \odot c \oplus rnd) =_q z\cdot m$
\end{theorem}
\begin{proof}
From the definition and equations, we expand the left hand side as follows:
\begin{equation}\label{eq:smuche_dec1}
\begin{split}
    &\; dec\left(z \odot c \oplus rnd \right)\\
=_q&\; dec\left( (z\cdot c_1 + PK_1\cdot \xi)_{\text{mod }q} , (z\cdot c_2 + PK_2\cdot \xi)_{\text{mod }q} \right)\\
=_q&\; dec((PK_1\cdot(z\cdot u + \xi) + z\cdot (e_1 + m) )_{\text{mod }q}, \\
&\;\hspace{7mm} (PK_2\cdot (z\cdot u + \xi) + z\cdot e_2)_{\text{mod } q})\\
=_q&\; dec\left( (PK_1\cdot u' + e'_1 + z\cdot m)_{\text{mod }q}, (PK_2\cdot u' + e'_2)_{\text{mod }q} \right),
\end{split}
\end{equation}
where $u' = z\cdot u + \xi$ and $e'_{1|2} = z\cdot e_{1|2}$. It should be noted that both $u'$ and $e'_{1|2}$ are still considered ``small'' polynomials in $z\cdot m$ because $u$, $\xi$, and $e_{1|2}$ are all considered ``small'' polynomials for $m$. Indeed, if we apply Eq.~\eqref{eq:ckks_dec}, the left hand side expands as follows:
\begin{equation}\label{eq:smuche_dec2}
\begin{split}
    &\; dec\left(z \odot c \oplus rnd \right) \hspace{6mm} \texttt{//continued from Eq.~\eqref{eq:smuche_dec1}}\\
=_q&\; PK_1\cdot u' + e'_1 + z\cdot m + SK\cdot(PK_2\cdot u' + e'_2)\\
=_q&\; -a\cdot SK\cdot u' + e'_1 + z\cdot m + a\cdot SK\cdot u' + SK\cdot e'_2\\
=_q&\; z\cdot m + (e'_1 + SK\cdot e'_2)\\
=_q&\; z\cdot m + \epsilon\\
=_q&\; z\cdot m,
\end{split}
\end{equation}
where $\epsilon = e'_1 + SK\cdot e'_2$ is a negligible term for $z\cdot m$.
\end{proof}

Back to our Smuche scheme, we simply apply Eq.~\eqref{eq:smuche_rnd} to the output of Alg.~\ref{alg:construct} to randomize the ciphertext.
That is, we output $c = c' \oplus rnd$ as the final ciphertext of $m$.
This randomization procedure takes a constant number of operations over the polynomials and is independent of the plaintext size.
We term the entire encryption algorithm $SmucheEnc$, following the naming convention of other state-of-the-art schemes $RacheEnc$ and $CkksEnc$.

\section{Security Analysis}

Only because $SmucheEnc$ is correct and randomized does not mean that it is secure. To demonstrate that Smuche is (semantically) secure, i.e., the ciphertext is indistinguishable from a random string even if an adversary obtains a polynomial number of plaintext-ciphertext pairs by arbitrarily choosing the plaintexts (IND-CPA), we need to reduce the breaking of $SmucheEnc$ output into an efficient algorithm for the underlying mathematical problem that is believed to be computationally infeasible to solve. In the context of modern HE schemes whose security is all based on the learning-with-error (LWE) problem, such a reduction would entail a probabilistic polynomial time (PPT) algorithm to solve the LWE problem. However, in this work we will assume the underlying HE $enc$ is IND-CPA and we will attempt to reduce breaking $SmucheEnc$ to breaking $enc$.

Formally, we will prove the following theorem.
\begin{theorem}
    Smuche is IND-CPA secure as long as the underlying HE encryption is IND-CPA secure.
\end{theorem}
\begin{proof}
Suppose $SmucheEnc$ is not IND-CPA secure but the underlying HE's $enc$ is INC-DPA secure. Then there exists an algorithm $\mathcal{A}$ such that:
\begin{itemize}
    \item $\mathcal{A}$ chooses two distinct plaintext messages $m^0$ and $m^1$;\footnote{The literature usually use subscript to denote two distinct messages; here we use superscript for this purpose because subscripts are reserved for the multiple polynomials inside a single ciphertext.}
    \item $\mathcal{A}$ can make a polynomial number of queries for the ciphertext of an arbitrary plaintext message (including $m^0$ and $m^1$);
    \item $Smuche$ randomly picks $i \in \{0,1\}$ and return $SmucheEnc(m^i)$ to $\mathcal{A}$;
    \item $\mathcal{A}$ can make the correct decision about $i \in \{0, 1\}$ with non-negligible advantage such that $dec(c) = m^i$.
\end{itemize}
On the other hand, we assume that no such algorithm exists for the underlying HE's encryption algorithm $enc$. Our goal here is to derive a contradiction to the above assumption.

If $\mathcal{A}$ can make the correct guess of $i$ with a non-negligible advantage, let $\frac{1}{poly(n)}$ denote the probability where $n$ denotes the security parameter.
There are two possible cases regarding Eq.~\eqref{eq:smuche_rnd}:
\begin{enumerate}
    \item If $rnd$ is repeated in the queries of $\mathcal{A}$ and in $SmucheEnc$, then $\mathcal{A}$ can immediately guess the correct $i$ for $z\cdot enc(m^i)$. Because $z$ is a constant, this entails that $\mathcal{A}$ can efficiently distinguish $enc(m^i)$ from a random string. The advantage for this case is $\frac{poly(n)}{2^n}$. Note that in this case, $\mathcal{A}$ can make the distinction for certain.
    
    \item If $rnd$ is not repeated in the polynomial number of queries as in $SmucheEnc$, this means that $\mathcal{A}$ can still make the correct guess with an advantage of $\displaystyle \gamma(n) = \frac{1}{poly(n)} - \frac{poly(n)}{2^n}$, which is not a negligible function in $n$. In other words, $\mathcal{A}$ can make a correct guess of $i$ with a non-negligible advantage $\gamma(n)$ such that $c = SmucheEnc(m^i)$ without ever seeing the ciphertext before. Note that $\mathcal{A}$ could succeed no matter whatever value is applied to $z\cdot c$, including zeros. Let $\mathcal{B}$ be such an algorithm where $rnd = (0, 0)$. As in the first case, $z$ is a constant, meaning that $\mathcal{B}$ can efficiently guess $i \in \{0, 1\}$ such that $c = enc(m^i)$, with the advantage of $\gamma(n)$. Also note that the $\mathcal{B}$ can be simulated by $\mathcal{A}$ in constant time, meaning that $\mathcal{B}$ is also a PPT algorithm.
    However, we assume that the underlying HE $enc$ is IND-CPA secure; therefore, such an algorithm $\mathcal{B}$ cannot exist, leading to a contradiction.
\end{enumerate}
\end{proof}

\section{Evaluation}

\subsection{System Implementation}

We have implemented the proposed Smuche caching scheme from scratch along with baseline prototypes of Rache~\cite{otawose_sigmod23} and CKKS~\cite{ckks}. Our implementation comprises about 3,000 lines of Java code. The project is compiled and executed with OpenJDK 21.0.1. We will publish the implementation as an open-source project:
\url{https://github.com/hpdic/smuche}.

The baseline HE scheme we choose to use is CKKS~\cite{ckks}, which is based on arithmetic circuits. There are a few important parameters we must discuss.
\begin{itemize}
    \item Slot. CKKS supports a more general set of plaintext messages than scalar floats: CKKS supports a vector of complex numbers. However, in our experiments, the workloads only consist of scalar floats. As a result, we set the number of slots in CKKS as 1. Inside the CKKS encoding procedure, however, there is a technical requirement to extend the original values with their complex conjugates. Therefore, although we set the slot to 1, what CKKS does is encode the single slot into two placeholders. We will not go into the details of this and we refer interested readers to the original CKKS paper~\cite{ckks}.

    \item Level. CKKS is one of the so-called leveled HE schemes, meaning that there is an upper bound for the number of multiplications between a pair of ciphertext messages. The reason why ciphertext multiplication $\otimes$ matters is that the inherent noise increase is quadratic. However, this is not a concern for scalar multiplication $\odot$, which is essentially a series of ciphertext addition $\oplus$. In our implementation, the level is set to 10. In production systems, this value is usually set to a much larger value such that the ciphertexts remain valid to decrypt. An alternative approach is bootstrapping, which resets the noise level; however, bootstrapping is rarely used in practice due to the overwhelming computational overhead.

    \item Precision. This metric refers to the number of digits to be considered accurate. Smuche supports both integers and float numbers, so there are two precision levels for the integer and fractional parts, respectively. We set both precision levels to 10 in our experiments.
\end{itemize}

It should be clear that the proposed Smuche is not bound to a specific HE scheme, although we choose to implement it on top of CKKS. The primary reason for our decision is that CKKS is the only scheme that supports efficient homomorphic operations on float numbers. Smuche can be built on top of all HE schemes that are based on arithmetic circuits: Recall that Smuche only manipulates a pair of polynomials in the ciphertext message in addition to the homomorphic addition operation of the underlying HE scheme.

\subsection{Experimental Setup}

For our system prototype deployment, we utilize the Chameleon Cloud infrastructure~\cite{katek_atc20}. Our assigned node offers the following specifications.
CPUs: Two Intel Gold 6240R CPUs running at a frequency of 2.40 GHz, providing a total of 96 threads.
Storage: A 480 GB Micron SATA SSD.
RAM: 192 GB of memory.
The operating system image is Ubuntu 22.04 LTS. 

In addition to microbenchmarks where we simply encrypt random numbers, we will evaluate the proposed Smuche scheme on three real-world data sets.
The first real-world application is the U.S. national Covid-19 statistics from April 2020 to March 2021~\cite{covid19data}.
The data set has 341 days of 16 metrics, such as \textit{death increase}, \textit{positive increase}, and \textit{hospitalized increase}.
The second real-world application is the history of Bitcoin trade volume~\cite{bitcoin_trade} since it was first exchanged in the public in February 2013.
The data consists of the accumulated Bitcoin exchange on a 3-day basis from February 2013 to January 2022,
totaling 1,086 floating-point numbers.
The third real-world application is the human genome reference 38~\cite{hg_data},
commonly known as \textit{hg38},
which includes 34,424 rows of singular attributes,
e.g., \textit{transcription positions}, \textit{coding regions}, and \textit{number of exons}, last updated in March 2020.

Some important parameters we use are as follows. The radix value $r$ is set to 2 in our experiments. The number of pivots, i.e., the number of cached ciphertexts, is denoted by $nPivot$, which ranges between 5 and 30. Unless otherwise stated, 40 integers are encrypted by different schemes. 

We did not turn on any parallel processing features such as OpenMP~\cite{openmp}. All experiments are repeated at least five times and we report the average numbers.

\subsection{Scalability Limitation of Rache}

Our first experiment demonstrates the scalability issues with Rache~\cite{otawose_sigmod23}, which is important as it is the motivation of this paper. In particular, we will compare the computational cost of Rache and CKKS when a variety number of pivots are employed. Our Java implementation of both CKKS and Rache is only a proof-of-concept and is far from optimal. Therefore, the raw performance is not comparable with popular frameworks such as HElib~\cite{helib} and SEAL~\cite{sealcrypto}. However, since all the baseline schemes and Smuche are built up on the same code base, the relative comparison among them remains valid.

\begin{table}
    \centering
    \begin{tabular}{lrr}
        \toprule
        $nPivot$  & $RacheEnc$ (ms) & Ratio over $CkksEnc$ \\
        \midrule
        4       & 0.74          & 0.43        \\
        8       & 1.04          & 0.59        \\
        16      & 1.46          & 0.83        \\
        32      & 2.09          & 1.21        \\
        64      & 3.65          & 2.10        \\
        \bottomrule
    \end{tabular}
    \caption{Scalability of Rache}
    \label{tbl:rache_scale}
\end{table}

Table~\ref{tbl:rache_scale} shows that when the number of pivots exceeds 32 the overhead of reconstructing the ciphertext outweighs the cost of the vanilla HE scheme, which makes it meaningless to apply Rache. The vanilla CKKS takes 1.74 ms to encrypt 40 numbers. It should be noted, however, that we did not turn on OpenMP in our implementation of Rache; therefore, in practice, the real threshold would be much larger than 32 before Rache becomes ineffective. The point is that the $RacheEnc$ time is proportional to the number of cached ciphertexts $nPivot$ and Rache is not applicable for a large plaintext space where a lot of cached ciphertexts are expected.  

\subsection{Computational Overhead of Smuche}

This experiment investigates the scalability of Smuche. Because Smuche only touches on one cached ciphertext, there is no point in scaling the number of pivots (as we did for Rache). Rather, we are more interested in whether the computational overhead is independent of the number of plaintext messages. This is called weak-scaling in the literature, where we scale the number of plaintext messages from 40 to 120.

\begin{table}
    \centering
    \begin{tabular}{lrr}
        \toprule
        \#Messages  & Overall Time (ms) & Time / Message (ms) \\
        \midrule
        40      & 0.87          & 0.02        \\
        60      & 1.22          & 0.02        \\
        80      & 1.54          & 0.02        \\
        100     & 1.81          & 0.02        \\
        120     & 1.99          & 0.02        \\
        \bottomrule
    \end{tabular}
    \caption{Encryption Performance of Smuche}
    \label{tbl:smuche_scale}
\end{table}

Table~\ref{tbl:smuche_scale} reports the encryption time of Smuche when a different number of plaintext messages are encrypted. We note that the overall encryption time is linearly proportional to the number of messages and the unit time to encrypt a single message remains constant, i.e., 0.02 ms. This result confirms the claimed property of Smuche: The per-message encryption performance is independent of the caching parameters such as the number of cached pivots or the number of messages.

\subsection{Real-World Data Sets}

We compare the end-to-end encryption performance of three HE schemes (CKKS, Rache, Smuche) with three real-world data sets (Covid-19, Bitcoin, Human Gene \#38). We set the number of pivots as 32 (i.e., $nPivot = 32$ for Bitcoin and a smaller one $nPivot = 16$ to Covid-19 and Human Gene \#38 because the trade volume in the Bitcoin data set is very large, i.e., up to one trillion. It should be noted that such a large $nPivot$ will like degrade Rache into even lower performance than the vanilla CKKS.

\begin{table}
    \centering
    \begin{tabular}{lrrr}
        \toprule
                  & Covid-19      & Bitcoin       & Human Gene \#38 \\
        \midrule
        CKKS    & 7.90 ms         & 20.76 ms      & 457.80 ms\\
        Rache   & 7.33 ms         & 29.64 ms      & 325.48 ms\\
        Smuche  & 2.74 ms         & 8.09 ms       & 225.99 ms\\
        Speedup  & 2.68$\times$   & 2.57$\times$ & 1.44$\times$ \\
        \bottomrule
    \end{tabular}
    \caption{Performance Comparison on Real-World Data Sets}
    \label{tbl:smuche_apps}
\end{table}

Table~\ref{tbl:smuche_apps} reports the performance of those HE schemes on our real-world data sets.
As expected, Rache does exhibit some performance degradation due to the large number of pivots on the Bitcoin data set. 
However, Smuche shows a clear advantage over both CKKS and Rache on all of the three data sets. 
The speed up of Smuche ranges between 1.44 (Human Gene \#38) and 2.68 (Covid-19).

\section{Conclusion}

In conclusion, this paper presents Smuche, a constant-time caching technique designed to bolster the efficiency of homomorphic encryption (HE) in secure machine learning for large-scale data. The provable security of Smuche is established through a reduction framework linked to the IND-CPA security of the underlying HE scheme. Comparative experiments with Rache and CKKS affirm the effectiveness of Smuche, showcasing its ability to achieve constant-time caching with markedly reduced overhead. The findings emphasize Smuche's practical utility, contributing insights into the ongoing challenge of optimizing the delicate balance between security and performance in the deployment of machine learning systems.

\bibliographystyle{named}
\bibliography{ijcai24}

\end{document}